\documentstyle[preprint,aps]{revtex}

\begin{document}
\draft
\title{Poincar\'{e} gauge theory of (2+1)-dimensional gravity}
\author{Toshiharu Kawai\thanks{
Electronic address: h1758@ocugw.cc.osaka-cu.ac.jp}}
\address{Department of Physics, Osaka City University, 3-3-138
Sugimoto, Sumiyoshi-ku, Osaka 558, Japan}
\maketitle
\begin{abstract}
A Poincar\'{e} gauge theory of (2+1)-dimensional gravity is developed.
Fundamental gravitational field variables are dreibein fields and
Lorentz gauge potentials, and the theory is
underlain with the Riemann-Cartan space-time. The most general
gravitational Lagrangian density, which is at most quadratic in
curvature and torsion tensors and invariant under local Lorentz
transformations and under general coordinate transformations, is given.
Gravitational field equations are studied in detail, and
solutions of the equations for weak gravitational fields are examined
for the case with a static, \lq \lq spin"less point like source.
We find, among other things, the following: (1)Solutions of the
vacuum Einstein equation satisfy gravitational field equations in
the vacuum in this theory. (2)For a class of the parameters
in the gravitational Lagrangian density, the torsion is
\lq \lq frozen" at the place where \lq \lq spin"
density of the source field is not vanishing. In this case, the field
equation actually agrees with the Einstein equation, when the source
field is \lq \lq spin"less. (3)A teleparallel theory developed in a
previous paper is \lq \lq included as a solution" in a limiting case.
(4)A Newtonian limit is obtainable, if the parameters in the
Lagrangian density satisfy certain conditions.
\end{abstract}
\narrowtext
\section{INTRODUCTION}
			    \setcounter{equation}{0}
Recently, lower dimensional gravity has been attracting
considerable attentions. The (2+1)-dimensional Einstein theory has
no Newtonian limit and no dynamical degrees of freedom, but it has
non-trivial global structures. This theory has been studied mainly
because of the local triviality and of the global non-triviality
[1-3]. For the (3+1)-dimensional gravity, there have been
proposed various theories alternative to the Einstein theory,
among which we have a teleparallel theory [4] and Poincar\'{e}
gauge theories [5--9]. A Poincar\'{e} gauge theory has been examined
[10] also for the (1+1)-dimensional case.

It would be significant to develop various theories of gravity
also for (2+1)-dimensional case, which will bring us toy
models useful to examine basic concepts in theories of gravity.
In a previous paper [11], the present author has proposed a
teleparallel theory of (2+1)-dimensional gravity having a
Newtonian limit and black hole solutions.

The purpose of this paper is to develop a Poincar\'{e} gauge theory
of (2+1)-dimensional gravity, in a limiting case of which a
teleparallel theory given in Ref. [11] is \lq \lq included as a
solution".
\section{DREIBEINS, THREE-DIMENSIONAL LORENTZ GAUGE POTENTIAL
AND RIEMANN-CARTAN SPACE-TIME}
			    \setcounter{equation}{0}

The three-dimensional space-time $M$ is assumed to be a
differentiable manifold endowed with the Lorentzian metric
$g_{\mu \nu }dx^{\mu }\otimes dx^{\nu }$ $(\mu ,\nu =0,1,2)$
related to the fields ${\bf e}^{k}={e^{k}}_{\mu }dx^{\mu }$
$(k=0,1,2)$ through the relation $g_{\mu \nu }=
{e^{k}}_{\mu }\eta_{kl}{e^{l}}_{\nu }$
with $(\eta_{kl}) \stackrel{\rm def}{=} diag(-1,1,1)$. Here,
$\{ x^{\mu };$ $\mu $ $ =0,1,2\}$ is a local coordinate of the
space-time. The fields ${\bf e}_{k}={e^{\mu}}_{k}\partial
/\partial x^{\mu }$, which  are dual to ${\bf e}^{k}$,
are the dreibein fields.
The Lorentz gauge potentials ${A^{kl}}_{\mu }$ $(=-{A^{lk}}_{\mu })$
transform according as
\begin{eqnarray}
      {A'^{kl}}_{\mu }(x) & = & {A^{kl}}_{\mu }(x)+{\omega^k}_{m}(x)
                                {A^{ml}}_{\mu }(x) \nonumber \\
                          & & +{\omega^{l}}_{m}(x){A^{km}}_{\mu }(x)
                              -\partial _{\mu }\omega^{kl}(x)\;,
\end{eqnarray}
under the infinitesimal Lorentz gauge transformation of
${e^{k}}_{\mu }(x)$:
\begin{equation}
      {e'^{k}}_{\mu }(x)={e^{k}}_{\mu }(x)
                         +{\omega^{k}}_{l}(x){e^{l}}_{\mu }(x)\;,
\end{equation}
where ${\omega^{kl}} \stackrel{\rm def}{=} \eta^{lm}{\omega^k}_{m}$
is an infinitesimal real valued function of $x$ and is antisymmetric
with respect to $k$ and $l$. Here, $(\eta^{kl})$ is the inverse
matrix of $(\eta_{kl})$, and in what follows, raising and lowering
the indices {\em k}, {\em l}, {\em m},..... are accomplished with
the aid of $(\eta^{kl})$ and $(\eta_{kl)}$.
The covariant derivative $D_{k}\varphi$ of the field $\varphi $
belonging to a representation $\sigma$ of the three-dimensional
Lorentz group is given by
\begin{equation}
     D_{k}\varphi ={e^{\mu }}_{k}\left(\partial _{\mu}\varphi
                   +\frac{i}{2}{A^{lm}}_{\mu }M_{lm}\varphi \right)\;,
\end{equation}
where $M_{kl} \stackrel{\rm def}{=} -i\sigma _{*}
({\overline{M}}_{kl})$. Here, ${\{\overline{M}}_{kl},k,l=0,1,2\}$
is a basis of the Lie algebra of the three-dimensional Lorentz group
satisfying the relation,
\begin{eqnarray}
    [{\overline{M}}_{kl}, {\overline{M}}_{mn}]
   & = & -\eta_{km}{\overline{M}}_{ln}
       -\eta_{ln}{\overline{M}}_{km} \nonumber \\
   & & +\eta_{kn}{\overline{M}}_{lm}
       +\eta_{lm}{\overline{M}}_{kn}\;,
\end{eqnarray}
\begin{equation}
    {\overline{M}}_{kl}=-{\overline{M}_{lk}}\;,
\end{equation}
and $\sigma _{*}$ stands for the differential
of $\sigma $. The field strengths of ${e^{k}}_{\mu }$ and of
${A^{kl}}_{\mu }$ are given by
\begin{eqnarray}
     {T^{k}}_{lm} \stackrel{\rm def}{=} {e^{\mu }}_{l}{e^{\nu }}_{m}
                   (\partial_{\mu }{e^{k}}_{\nu }-\partial_{\nu }
                   {e^{k}}_{\mu }) \nonumber \\
                   +{e^{\mu }}_{l}{A^{k}}_{m\mu }
                   -{e^{\mu }}_{m}{A^{k}}_{l\mu }\;,
\end{eqnarray}
\begin{eqnarray}
     {R^{k}}_{lmn} \stackrel{\rm def}{=} {e^{\mu }}_{m}{e^{\nu }}_{n}
                    (\partial_{\mu }{A^{kl}}_{\nu }-\partial_{\nu }
                    {A^{kl}}_{\mu } \nonumber \\
                    -{A^{k}}_{r\mu }{A^{lr}}_{\nu }
                    +{A^{k}}_{r\nu }{A^{lr}}_{\mu })\;,
\end{eqnarray}
respectively.
We have the relation,
\begin{eqnarray}
      R_{klmn} &= &\eta_{km}R_{ln}-\eta_{kn}R_{lm}-\eta_{lm}R_{kn}+
               \eta_{ln}R_{km} \nonumber \\
          & &  -\frac{1}{2}(\eta_{km}\eta _{ln}
               -\eta_{kn}\eta_{lm})R\;,
\end{eqnarray}
where we have defined $R_{kl} \stackrel{\rm def}{=}
{{R_{k}}^{m}}_{lm}$ and $R \stackrel{\rm def}{=} {R^{k}}_{k}$.

For the world vector field ${\bf V}=V^{\mu }\partial/
{\partial x^{\mu }}$,
the covariant derivative with respect to the affine connection
${\Gamma}^{\mu }_{\lambda \nu }$ is given by
\begin{equation}
     D_{\nu }V^{\mu }=\partial_{\nu }V^{\mu }+\Gamma^{\mu }_{\lambda
                      \nu }V^{\lambda }\;.
\end{equation}
     We make the requirement,
\begin{equation}
     D_{l}V^{k}={e^{\nu }}_{l}{e^{k}}_{\mu }D_{\nu }V^{\mu }\;,
\end{equation}
for the Lorentz vector field $V^{k}$, where $V^{\mu }
\stackrel{\rm def}{=} {e^{\mu }}_{k}V^{k}$. Then the relation,
\begin{equation}
     {A^{k}}_{l\mu }\equiv\Gamma ^{\nu }_{\lambda \mu }{e^{k}}_
                     {\nu }{e^{\lambda }}_{l}+{e^{k}}_{\nu }
                     \partial_{\mu }{e^{\nu }}_{l}\;,
\end{equation}
follows, and we have
\begin{equation}
     {T^{k}}_{\mu \nu }\equiv{e^{k}}_{\lambda }
                        {T^{\lambda }}_{\mu \nu }\;,
\end{equation}
\begin{equation}
     {R^{k}}_{l \mu \nu }\equiv{e^{k}}_{\lambda }
                          {e^{\rho }}_{l}{R^{\lambda }}_
                          {\rho \mu \nu }\;,
\end{equation}
\begin{equation}
      {D_{\lambda }}g_{\mu \nu } \stackrel{\rm def}{=} \partial_
                                  {\lambda }g_{\mu \nu }-
                                  \Gamma^{\rho }_{\mu \lambda }
                                  g_{\rho \nu }-\Gamma^{\rho }_
                                 {\nu \lambda }g_{\mu \rho} \equiv 0
\end{equation}
with
\begin{equation}
      {T^{\mu }}_{\nu \lambda } \stackrel{\rm def}{=} \Gamma^{\mu}_
                                 {\lambda \nu }-\Gamma^{\mu }_{\nu
                                 \lambda }\;,
\end{equation}
\begin{equation}
      {R^{\mu }}_{\nu \lambda \rho } \stackrel{\rm def}{=} \partial_
                                      {\lambda }\Gamma^{\mu }_
                                      {\nu \rho }-\partial_{\rho }
                                      \Gamma^{\mu }_{\nu \lambda }
                                      +\Gamma^{\mu }_{\tau \lambda }
                                      \Gamma^{\tau }_{\nu \rho }-
                                      \Gamma^{\mu }_{\tau \rho }
                                      \Gamma^{\tau }_{\nu \lambda }\;.
\end{equation}
The components ${T^{\mu }}_{\nu \lambda }$
and ${R^{\mu }}_
{\nu \lambda \rho }$ are those of the torsion tensor and of
the curvature tensor, respectively, and they are both
non-vanishing in general. Thus, the space-time $M$ is of the
Riemann-Cartan type. From (2.14), we obtain
\begin{equation}
     \Gamma^{\lambda }_{\mu \nu}=
 \mbox{\tiny{$\left\{\mbox{\hspace*{-1.5ex}}\begin{array}{c}
     \mbox{\scriptsize $\lambda $}\\ \mbox{\scriptsize $\mu \: \nu $}
     \end{array}\mbox{\hspace*{-1.5ex}}\right\}$}}
     +{K^{\lambda}}_{\mu \nu }\;,
\end{equation}
where the first term denotes the Christoffel symbol,
\begin{equation}
 \mbox{\tiny{$\left\{\mbox{\hspace*{-1.5ex}}\begin{array}{c}
     \mbox{\scriptsize $\lambda $}\\ \mbox{\scriptsize $\mu \: \nu $}
     \end{array}\mbox{\hspace*{-1.5ex}}\right\}$}}
     \stackrel{\rm def}{=} \frac{1}{2}g^{\lambda \xi}
                           (\partial_{\mu }g_{\xi \nu }
                           +\partial_{\nu}g_{\xi \mu }-
                           \partial_{\xi }g_{\mu \nu})\;,
\end{equation}
and the second stands for the contortion tensor,
\begin{equation}
     {K^{\lambda}}_{\mu \nu } \stackrel{\rm def}{=} -\frac{1}{2}
                              \left({T^{\lambda }}_{\mu \nu}-
     {{T_{\mu }}^{\lambda}}_{\raisebox{.4ex}{\scriptsize$\nu $}}-
     {{T_{\nu }}^{\lambda}}_{\raisebox{.2ex}{\scriptsize $\mu $}}\right)
     \;.
\end{equation}
The field components ${e^{k}}_{\mu }$ and ${e^{\mu }}_{k}$ will be
used, as for the case of $V^{\mu }$ and $V^{k}$ in the above,
to convert Latin and Greek indices.
\section{LAGRANGIAN DENSITIES AND GRAVITATIONAL FIELD EQUATIONS}
			    \setcounter{equation}{0}
For the matter field $\varphi $, $L_{M}(\varphi ,D_{k}\varphi )$
is a Lag\-ran\-gian [12] invariant under three-di\-men\-sion\-al local
Lorentz transformations and under general coordinate
transformations, if $L_{M}(\varphi , \partial_{k}\varphi )$ is an
invariant Lag\-ran\-gian on the three-di\-men\-sion\-al Min\-kow\-ski
space-time.

For the fields ${e^{k}}_{\mu }$ and ${A^{kl}}_{\mu }$, Lagrangians,
which are invariant under local Lorentz transformations including
also inversions and under general coordinate transformations and at
most quadratic in torsion and curvature tensors, are given by
\begin{equation}
      L_{T}=\alpha t^{klm}t_{klm}+\beta v^{k}v_{k}+\gamma
      a^{klm}a_{klm}+\delta \;,
\end{equation}
\begin{equation}
      L_{R}=a_{1}E^{kl}E_{kl}+a_{2}I^{kl}I_{kl}+a_{3}R^{2}+aR\;.
\end{equation}
Here, $t_{klm}, v_{k}$ and $a_{klm}$ are the irreducible components
of $T_{klm}$ defined by
\begin{equation}
      t_{klm} \stackrel{\rm def}{=} \frac{1}{2}(T_{klm}+T_{lkm})
               +\frac{1}{4}(\eta_{mk}v_{l}+\eta_{ml}v_{k})-
               \frac{1}{2}\eta_{kl}v_{m}\;,
\end{equation}
\begin{equation}
      v_{k} \stackrel{\rm def}{=} {T^{l}}_{lk}\;,
\end{equation}
and
\begin{equation}
      a_{klm} \stackrel{\rm def}{=} \frac{1}{3}(T_{klm}+
               T_{mkl}+T_{lmk})\;,
\end{equation}
respectively, and $E_{kl}$ and $I_{kl}$
are the irreducible components of $R_{klmn}$ defined by
\begin{equation}
      E_{kl} \stackrel{\rm def}{=} \frac{1}{2}(R_{kl}-R_{lk})
\end{equation}
and
\begin{equation}
      I_{kl} \stackrel{\rm def}{=} \frac{1}{2}(R_{kl}+R_{lk})-
      \frac{1}{3}\eta_{kl}R\;,
\end{equation}
respectively. Also, $\alpha $, $\beta $, $\gamma $,
$\delta $, $a_{1}$, $a_{2}$, $a_{3}$ and $a$ are real constant
parameters. Then,
\begin{equation}
      {\bf I} \stackrel{\rm def}{=} \frac{1}{c}\int
                                    {\bf L}d^{3}x
\end{equation}
is the total action of the system, where $c$ is the light velocity
in the vacuum and ${\bf L}$ is defined by
\begin{equation}
      {\bf L} \stackrel{\rm def}{=}\sqrt{-g}(L_{G}+
                           L_{M}(\varphi ,D_{k}\varphi ))
\end{equation}
with $L_{G} \stackrel{\rm def}{=} L_{T}+L_{R}$ and
$g \stackrel{\rm def}{=} det(g_{\mu \nu})$.
For the case with $a_{1}=a_{2}=a_{3}=0$, $a \neq 0$, $\alpha =\beta
=\gamma =0$ and with $\delta =0$, the Lagrangian $L_{G}$ reduces
to the Einstein-Cartan Lagrangian [13, 14].
The field equation ${\delta {\bf L}}/
{\delta {e^{i}}_{\mu }}=0$ reads [15]
\begin{eqnarray}
     2aR_{ji}+4J_{[ik][jl]}R^{kl}+4{J^{[kl]}}_{[jl]}R_{ki}-
     2{{J_{[i}}^{k]}}_{[jk]}R \nonumber \\
     -2D^{k}F_{ijk}+2v^{k}F_{ijk}+2H_{ij}-\eta_{ij}L_{G}=T_{ij}\;,
\end{eqnarray}
where we have defined
\begin{equation}
     J_{ijkl} \stackrel{\rm def}{=}2a_{3}R\eta_{ik}\eta_{jl}
               +2\eta_{ik}(a_{1}E_{jl}+a_{2}I_{jl})\;,
\end{equation}
\begin{eqnarray}
     D^{k}F_{ijk} & \stackrel{\rm def}{=} & e^{\mu k}(\partial_{\mu }
                   F_{ijk}+{{A_{i}}^{m}}_{\mu }F_{mjk} \nonumber \\
              & &  +{{A_{j}}^{m}}_{\mu }F_{imk}
                   +{{A_{k}}^{m}}_{\mu }F_{ijm})\;,
\end{eqnarray}
\begin{eqnarray}
     F_{ijk} & \stackrel{\rm def}{=} & \alpha (t_{ijk}-t_{ikj})
               +\beta (\eta_{ij}v_{k}-\eta_{ik}v_{j})
              +2\gamma a_{ijk} \nonumber \\
             &=&-F_{ikj}
\end{eqnarray}
and
\begin{equation}
     H_{ij} \stackrel{\rm def}{=} T_{kli}{F^{kl}}_{j}-
             \frac{1}{2}T_{jkl}{F_{i}}^{kl}=H_{ji}\;.
\end{equation}
Also, $T_{ij}$ denotes the energy-momentum density of the field
$\varphi $ defined by
\begin{equation}
     \sqrt{-g}T_{ij} \stackrel{\rm def}{=} e_{j\mu }\frac{\delta
                      (\sqrt{-g}L_{M})}{\delta {e^{i}}_{\mu }}\;.
\end{equation}
The field equation ${\delta {\bf L}}/
{\delta {A^{ij}}_{\mu }}=0$ reads
\begin{eqnarray}
     2D^{l}J_{[ij][kl]}+\left(\frac{4}{3}{t_{k}}^{[lm]}-
     {\delta_{k}}^{[l}v^{m]}
     +{a_{k}}^{lm}\right)J_{[ij][lm]} \nonumber \\
     -H_{ijk}=S_{ijk}
\end{eqnarray}
with
\begin{eqnarray}
     D^{l}J_{[ij][kl]} & \stackrel{\rm def}{=} & e^{\mu l}
                        (\partial_{\mu}J_{[ij][kl]}
                        +{{A_{i}}^{m}}_{\mu }
                        J_{[mj][kl]} \nonumber \\
                   & &  +{{A_{j}}^{m}}_{\mu }J_{[im][kl]}
                        +{{A_{k}}^{m}}_{\mu }J_{[ij][ml]} \nonumber \\
                   & &  +{{A_{l}}^{m}}_{\mu }J_{[ij][km]})\;,
\end{eqnarray}
\begin{eqnarray}
\lefteqn{H_{ijk} \stackrel{\rm def}{=}
          -\left(\alpha +\frac{2a}{3}\right)
          (t_{kij}-t_{kji})
          -\left(\beta -\frac{a}{2} \right)} \nonumber \\
          & & \times (\eta_{ki}v_{j}-\eta_{kj}v_{i})+(4\gamma -a)
              a_{ijk}=-H_{jik}\;.
\end{eqnarray}
Here, $S_{ijk}$ is the "spin" [16] density of $\varphi $ defined by
\begin{equation}
     \sqrt{-g}S_{ijk} \stackrel{\rm def}{=} -e_{k\mu }\frac{\delta
                       (\sqrt{-g}L_{M})}{\delta {A^{ij}}_{\mu }}\;.
\end{equation}
\section{ALTERNATIVE FORMS OF THE GRAVITATIONAL FIELD EQUATIONS}
			    \setcounter{equation}{0}
We shall rewrite the gravitational field equations (3.10) and (3.16),
by using the expression,
\begin{equation}
     A_{ij\mu}=\Delta_{ij\mu }+K_{ij\mu }
\end{equation}
with $\Delta_{ij\mu }$ being the Ricci rotation coefficient,
\begin{equation}
     \Delta_{ij\mu } \stackrel{\rm def}{=} \frac{1}{2}{e^{k}}_{\mu }
                      (C_{ijk}-C_{jik}-C_{kij})\;,
\end{equation}
where
\begin{equation}
     C_{ijk} \stackrel{\rm def}{=} {e^{\nu }}_{j}{e^{\lambda }}_{k}
              (\partial_{\nu }e_{i\lambda }-\partial_{\lambda }
              e_{i\nu })\;.
\end{equation}
There is the relation,
\begin{equation}
      R_{ij\mu \nu }=R_{ij\mu \nu }(\{\})+R_{ij\mu \nu}(K)
\end{equation}
with
\begin{eqnarray}
      R_{ij\mu \nu }(\{\}) & \stackrel{\rm def}{=} & \partial_{\mu }
                            \Delta_{ij\nu }-\partial_{\nu }\Delta_
                            {ij\mu }-{{\Delta_{i}}^{k}}_{\mu }
                            \Delta_{jk\nu }+{{\Delta_{i}}^{k}}_{\nu }
                            \Delta_{jk\mu } \nonumber \\
                         & = & e_{i\lambda }
                             {e^{\rho }}_{j}{R^{\lambda}}_{\rho \mu
                             \nu }(\{\})\;,
\end{eqnarray}
\begin{eqnarray}
      R_{ij\mu \nu }(K) & \stackrel{\rm def}{=} & \nabla_{\mu }
                          K_{ij\nu }
                         -\nabla_{\nu }K_{ij\mu }\nonumber \\
                    & &  -{{K_{i}}^{k}}_{\mu }
                         K_{jk\nu }+{{K_{i}}^{k}}_{\nu }K_{jk\mu }\;,
\end{eqnarray}
as is shown by substituting (4.1) into (2.7).
Here, ${R^{\lambda }}_{\rho \mu \nu }(\{\})$ stands for the
Riemann-Christoffel curvature tensor,
\begin{eqnarray}
      {R^{\lambda}}_{\rho \mu \nu }(\{\})
      & \stackrel{\rm def}{=} & \mbox{\small$ \partial$}_{\mu }
 \mbox{\tiny{$\left\{\mbox{\hspace*{-1.5ex}}\begin{array}{c}
     \mbox{\scriptsize $\lambda $}\\ \mbox{\scriptsize $\rho \: \nu $}
     \end{array}\mbox{\hspace*{-1.5ex}}\right\}$}}
        -\mbox{\small$ \partial$}_{\nu }
 \mbox{\tiny{$\left\{\mbox{\hspace*{-1.5ex}}\begin{array}{c}
     \mbox{\scriptsize $\lambda $}\\ \mbox{\scriptsize $\rho \: \mu $}
     \end{array}\mbox{\hspace*{-1.5ex}}\right\}$}}
     \nonumber \\
   & &
+\mbox{\tiny{$\left\{\mbox{\hspace*{-1.5ex}}\begin{array}{c}
     \mbox{\scriptsize $\lambda $}\\ \mbox{\scriptsize $\tau \: \mu $}
     \end{array}\mbox{\hspace*{-1.5ex}}\right\}$}}
 \mbox{\tiny{$\left\{\mbox{\hspace*{-1.5ex}}\begin{array}{c}
     \mbox{\scriptsize $\tau $}\\ \mbox{\scriptsize $\rho \: \nu $}
     \end{array}\mbox{\hspace*{-1.5ex}}\right\}$}}
 -\mbox{\tiny{$\left\{\mbox{\hspace*{-1.5ex}}\begin{array}{c}
     \mbox{\scriptsize $\lambda $}\\ \mbox{\scriptsize $\tau \: \nu $}
     \end{array}\mbox{\hspace*{-1.5ex}}\right\}$}}
 \mbox{\tiny{$\left\{\mbox{\hspace*{-1.5ex}}\begin{array}{c}
     \mbox{\scriptsize $\tau $}\\ \mbox{\scriptsize $\rho \: \mu $}
     \end{array}\mbox{\hspace*{-1.5ex}}\right\}$}}\;,
\end{eqnarray}
and $\nabla_{\mu }K_{ij\nu }$ denotes the covariant derivative
with respect to the Ricci rotation coefficients when the
index is Latin, and with respect to the Levi-Civita connection
when the index is Greek.
Each irreducible part of $R_{ijkl}$ is split into two parts,
as is known by using (4.4) in (3.6) and (3.7), and $J_{ijkl}$
can be expressed as
\begin{equation}
      J_{ijkl}=J_{ijkl}(\{\})+J_{ijkl}(K)\;,
\end{equation}
where $J_{ijkl}(\{\})$ and $J_{ijkl}(K)$ are formed of
the irreducible parts of $R_{ijkl}(\{\})$ and of $R_{ijkl}(K)$,
respectively. The tensor $J_{ijkl}(\{\})$, in particular, is given by
\begin{equation}
       J_{ijkl}(\{\})=2a_{2}\eta_{ik}R_{jl}(\{\})
                      +2\left(a_{3}-\frac{a_{2}}{3}\right)
                      \eta_{ik}\eta_{jl}R(\{\})\;,
\end{equation}
where $R_{ij}(\{\})$ and $R(\{\})$ are the Ricci tensor and the
Riemann-Christoffel scalar curvature, respectively,
\begin{equation}
       R_{ij}(\{\}) \stackrel{\rm def}{=} {e^{\mu }}_{i}
                     {e^{\nu }}_{j}{R^{\lambda }}_{\mu \lambda \nu }
                     (\{\})\;,
       R(\{\}) \stackrel{\rm def}{=} \eta^{ij}R_{ij}(\{\})\;.
\end{equation}
The gravitational Lagrangian $L_{G}$ can be rewritten as
\begin{eqnarray}
     L_{G} & = & a_{2}R^{kl}(\{\})R_{kl}(\{\})+\left(a_{3}-
             \frac{a_{2}}{3}\right)(R(\{\}))^{2}+aR(\{\})
             \nonumber \\
        & & +L'_{T}+L'_{R}-\frac{2a}{\sqrt{-g}}\partial_{\mu}
             (\sqrt{-g}v^{\mu })\;,
\end{eqnarray}
where
\begin{eqnarray}
       L'_{T} \stackrel{\rm def}{=} \left(\alpha +\frac{2a}{3}
               \right)t^{klm}t_{klm}+\left(\beta-\frac{a}{2}
               \right)v^{k}v_{k} \nonumber \\
               +\left(\gamma-\frac{a}{4}
               \right)a^{klm}a_{klm}+\delta \;,
\end{eqnarray}
\begin{eqnarray}
       L'_{R} & \stackrel{\rm def}{=} & L_{R}-a_{2}R^{kl}(\{\})
               R_{kl}(\{\}) \nonumber \\
           & & -\left(a_{3}-\frac{a_{2}}{3}\right)
               {(R(\{\}))}^{2}-aR\;.
\end{eqnarray}
Here, we have used the relation,
\begin{eqnarray}
       \sqrt{-g}R=\sqrt{-g}R(\{\})-\sqrt{-g}\left(-\frac{2}{3}
        t^{klm}t_{klm} \right. \nonumber \\
        \left. +\frac{1}{2}v^{k}v_{k}+\frac{1}{4}
        a^{klm}a_{klm}\right)
        -2\partial_{\mu }(\sqrt{-g}v^{\mu })\;.
\end{eqnarray}
Using the above formulae in (3.10), we get the alternative form of
the field equation for ${e^{i}}_{\mu }$,
\widetext
\begin{eqnarray}
  & & 2aG_{ij}(\{\})+\frac{1}{3}(5a_{2}+12a_{3})R_{ij}(\{\})R(\{\})
      -2a_{2}{R_{i}}^{k}(\{\})R_{jk}(\{\}) \nonumber \\
  & & +\eta_{ij}\left\{a_{2}R^{kl}(\{\})R_{kl}(\{\})
      -\left(\frac{2a_{2}}{3}+a_{3}\right)(R(\{\}))^2\right\}+a_{2}
      \eta_{ij}\{2R^{kl}(\{\})R_{kl}(K)-R(\{\})R(K)\} \nonumber \\
  & & +a_{2}R_{ij}(\{\})R(K)+\frac{2}{3}(a_{2}+6a_{3})R(\{\})
      R_{ji}(K) -2a_{2}{R_{i}}^{k}(\{\})R_{jk}(K)
      +4J_{[ik][jl]}(K)R^{kl} \nonumber \\
  & & +4{J^{[kl]}}_{[jl]}(K)R_{ki}-2{{J_{[i}}^{k]}}_{[jk]}(K)R
      -2D^{k}F'_{ijk}+2v^{k}F'_{ijk}+2H'_{ij} \nonumber \\
  & & -\eta_{ij}(L'_{T}+L'_{R})=T_{ij}\;,
\end{eqnarray}
\narrowtext
\noindent where $G_{ij}(\{\})$ is the three-dimensional Einstein
tensor,
\begin{equation}
      G_{ij}(\{\}) \stackrel{\rm def}{=} R_{ij}(\{\})-
              \frac{1}{2}\eta_{ij}R(\{\})\;.
\end{equation}
Here, we have defined
\begin{eqnarray}
      F'_{ijk} & \stackrel{\rm def}{=} &
                 \left(\mbox{\hspace*{-.6ex}}\alpha+
                \frac{2a}{3} \mbox{\hspace*{-.6ex}}\right)
                \mbox{\hspace*{-.3ex}}(t_{ijk}-t_{ikj})
                \mbox{\hspace*{-.3ex}}+\mbox{\hspace*{-.3ex}}
                \left(\beta-\frac{a}{2} \right)\mbox{\hspace*{-.3ex}}
                (\eta_{ij}v_{k}-\eta_{ik}v_{j}) \nonumber \\
            & & +2\left(\gamma-\frac{a}{4} \right)a_{ijk}=-F'_{ikj}\;,
\end{eqnarray}
\begin{equation}
      H'_{ij} \stackrel{\rm def}{=} T_{kli}{F'^{kl}}_{j}-
               \frac{1}{2}T_{jkl}{F'_{i}}^{kl}=H'_{ji}\;.
\end{equation}
We have the relations,
\begin{equation}
      F'_{ijk}=\frac{1}{2}(H_{ijk}-H_{ikj}-H_{jki})
\end{equation}
and
\begin{equation}
      H_{ijk}=F'_{ijk}-F'_{jik}\;,
\end{equation}
as is shown by comparing (3.18) and (4.17).
The field equation for ${A^{ij}}_{\mu }$ is rewritten as
\widetext
\begin{eqnarray}
      -2a_{2}\nabla_{[i}G_{j]k}(\{\})-8\left(a_{3}+
      \frac{a_{2}}{6}\right)\eta_{k[i}\partial_{j]}G(\{\})
      +2(D^{l}-\nabla^{l})J_{[ij][kl]}(\{\})
      +2D^{l}J_{[ij][kl]}(K) \nonumber \\
      +\left(\frac{4}{3}{t_{k}}^{[lm]}-{\delta_{k}}^{[l}v^{m]}
      +{a_{k}}^{lm} \right)J_{[ij][lm]}-H_{ijk}=S_{ijk}\;,
\end{eqnarray}
\narrowtext
\noindent where $G(\{\}) \stackrel{\rm def}{=}
\eta^{ij}G_{ij}(\{\})$. By examining the alternative forms of the
gravitational field equations (4.15) and (4.21), we find the
following:

(1) For the case with $S_{ijk}\equiv 0$ and with $T_{ij}\equiv 0$,
any solution of the equations
\begin{equation}
      G_{ij}(\{\})-\eta_{ij}\Lambda=0\;,
\end{equation}
\begin{equation}
      T_{ijk}=0\;,
\end{equation}
satisfies (4.15) and (4.21) with $\delta =2\Lambda (a+6a_{3}\Lambda )$.
We can say shortly, \lq \lq Solutions of the vacuum Einstein equation
are solutions of the vacuum gravitational field equations in this
theory."

(2) The equation (4.21) does not contain third derivatives of the
metric tensor, if and only if
\begin{equation}
      a_{2}=a_{3}=0\;.
\end{equation}

(3) When the condition (4.24) is satisfied, then (4.15) and (4.21)
are considerably simplified. In (4.15), the terms quadratic in the
Riemann-Christoffel curvature tensor are all vanishing. In (4.21),
the first three terms disappear, and all the remaining terms are
linear or quadratic in the torsion tensor. Thus, if the intrinsic
\lq \lq spin" of the source is vanishing, $S_{ijk}\equiv 0$, then
(4.21) is satisfied by the vanishing torsion, and when the torsion
vanishes, (4.15) reduces to the equation,
\begin{equation}
      2aG_{ij}(\{\})-\eta_{ij}\delta =T_{ij}\;.
\end{equation}
For the case with $a=1/2\kappa $ with $\kappa $ being the
\lq \lq Einstein gravitational constant", (4.25) agrees with the
Einstein equation, because $T_{ij}$=$T_{ji}$ for a vanishing $S_{ijk}$,
as is seen from (5.10) of the next section. The following,
however, should be noted:  {\em The torsion tensor
does not necessarily vanish, even when the condition} (4.24) {\em is
satisfied and the intrinsic \lq \lq spin" of the source field is
vanishing}.

(4) When the condition $a_{1}=0$ is satisfied in addition to the
condition (4.24), then (4.15) and (4.21) reduce to
\begin{equation}
    2aG_{ij}(\{\})-2D^{k}F'_{ijk}+2v^{k}F'_{ijk}+2H'_{ij}
    -\eta_{ij}L'_{T}=T_{ij}
\end{equation}
and
\begin{equation}
    -H_{ijk}=S_{ijk}\;,
\end{equation}
respectively. The torsion tensor is linearly dependent on $S_{ijk}$,
if
\begin{equation}
    (3\alpha+2a)(2\beta-a)(4\gamma-a) \neq 0\;,
\end{equation}
as is seen from (3.18) and (4.27). Thus, the torsion is
\lq \lq frozen" at the place where the
\lq \lq spin" density $S_{ijk}$ does not vanish.

If $S_{ijk}\equiv 0$, (4.15) for the present case reduces to (4.25).
Also, the field equations for the \lq \lq spin"less source fields
agree with those in the Einstein theory.
\section{EQUATION OF MOTION FOR MACROSCOPIC BODIES}
			    \setcounter{equation}{0}
We shall derive the equation of motion for macroscopic bodies,
which can be done in a way similar to the case of the
(3+1)-dimensional theory [7].

 From the fact that gravitational action integral,
\begin{equation}
      {\bf I}_{G} \stackrel{\rm def}{=} \frac{1}{c}\int
                                        {\bf L}_{G}d^{3}x
\end{equation}
with
\begin{equation}
      {\bf L}_{G} \stackrel{\rm def}{=} \sqrt{-g}L_{G}\;,
\end{equation}
is invariant under general coordinate transformations, the identity,
\begin{eqnarray}
       \sqrt{-g}&(&Y^{i\nu }\partial_{\mu }
       e_{i\nu }+{Z_{ij}}^{\nu }
       \partial_{\mu }{A^{ij}}_{\nu }) \nonumber \\
   & & \equiv \partial_{\nu }
       \{\sqrt{-g}({Y_{\mu }}^{\nu }
       +{Z_{ij}}^{\nu }{A^{ij}}_{\mu })\}\;,
\end{eqnarray}
follows, where
\begin{equation}
     \sqrt{-g}{Y_{i}}^{\mu } \stackrel{\rm def}{=}
                              -\frac{\delta{\bf L}_{G}}
                              {\delta {e^{i}}_{\mu }}
\end{equation}
and
\begin{equation}
     \sqrt{-g}{Z_{ij}}^{\mu } \stackrel{\rm def}{=}
                               -\frac{\delta {{\bf L}_{G}}}
                               {\delta {A^{ij}}_{\mu }}\;.
\end{equation}
The gravitational field equations ${\delta {\bf L}}/
{\delta {e^{i}}_{\mu }}=0$ and ${\delta {\bf L}}/
{\delta {A^{ij}}_{\mu }}=0$ can be expressed as
\begin{equation}
     {Y_{\mu }}^{\nu }={T_{\mu }}^{\nu }
\end{equation}
and
\begin{equation}
     {Z_{ij}}^{\nu }=-{S_{ij}}^{\nu }\;,
\end{equation}
respectively.
By using (5.6), (5.7) and the formula,
\begin{eqnarray}
     \partial_{\nu }&(&\sqrt{-g}{Y_{\mu }}^{\nu })
     -\sqrt{-g}Y^{i\nu } \partial_{\mu }e_{i\nu } \nonumber \\
 & & \equiv \sqrt{-g}(\nabla_{\nu}{Y_{\mu }}^{\nu }
     +\Delta_{\lambda \nu \mu }Y^{\lambda \nu })\;,
\end{eqnarray}
in (5.3), we obtain the response equation to gravitation,
\begin{eqnarray}
     \nabla_{\nu }{T_{\mu }}^{\nu }+\Delta_{\lambda \nu \mu }
     T^{\lambda \nu }-2\partial_{[\nu }{A^{ij}}_{\mu ]}
     {S_{ij}}^{\nu } \nonumber \\
     -\frac{1}{\sqrt{-g}}{A^{ij}}_{\mu }
     \partial_{\nu }(\sqrt{-g}{S_{ij}}^{\nu }) \equiv 0\;.
\end{eqnarray}
Also, the identity,
\begin{equation}
     \sqrt{-g}T_{[ij]}\equiv \nabla_{\mu }(\sqrt{-g}{S_{ij}}^{\mu })
                       -\frac{i}{2}\frac{\delta {\bf L}_{M}}
                       {\delta \varphi }M_{ij}\varphi \;,
\end{equation}
follows from the invariance of the total action ${\bf I}$
under local Lorentz transformations. Here, we have defined
\begin{eqnarray}
        \nabla_{\mu }(\sqrt{-g}{S_{ij}}^{\mu })
        & \stackrel{\rm def}{=} &
         \partial_{\mu }(\sqrt{-g}{S_{ij}}^{\mu })
         +\sqrt{-g}{{A_{i}}^{m}}_{\mu }{S_{mj}}^{\mu } \nonumber \\
     & & +\sqrt{-g}{{A_{j}}^{m}}_{\mu }{S_{im}}^{\mu }\;.
\end{eqnarray}
We obtain
\begin{eqnarray}
   \nabla_{\nu }{T_{\mu }}^{\nu }+ & \frac{1}{\sqrt{-g}} &
   \Delta_{ij\mu }\nabla_{\nu }(\sqrt{-g}S^{ij\nu })
   -2\partial_{[\nu }{A^{ij}}_{\mu ]}{S_{ij}}^{\nu } \nonumber \\
   & & -\frac{1}{\sqrt{-g}}{A^{ij}}_{\mu }
       \partial_{\nu }(\sqrt{-g}{S_{ij}}^{\nu })=0\;,
\end{eqnarray}
by the use of (5.10) and of the field equation
$\delta {\bf L}/\delta \varphi =\delta {\bf L}_{M}/
{\delta \varphi }=0$ in (5.9). We apply this equation to the motion of
a macroscopic body for which effects due to the \lq \lq spin"s of
the constituent fundamental particles can be ignored, then the
energy-momentum tensor of a macroscopic body is symmetric
and satisfies the conservation laws,
\begin{equation}
   \nabla_{\nu }{T_{\mu }}^{\nu }=0\;.
\end{equation}
 From (5.13), we can show, in a way quite similar to the case in the
four-dimensional Einstein theory, that the world line of a macroscopic
body is the geodesic line of the metric $g_{\mu \nu }
dx^{\mu }\otimes dx^{\nu }$.
\section{TWO LIMITING CASES}
			    \setcounter{equation}{0}
\subsection{The case with $a_{i} \rightarrow \infty (i=1,2,3)$}
Suppose that the parameters $a_{i} (i=1,2,3)$ have the expression
\begin{equation}
a_{i}=\frac{1}{f^{2}}{\overline{a}}_{i}, \;(i=1,2,3)\;,
\end{equation}
where $f$ is a parameter characterizing the magnitude of $a_{i}$'s
and it can be regarded as standing for the coupling strength between
the Lorentz gauge field and the matter field $\varphi$. Multiplying
both sides of (3.16) by $f^2$ and taking the limit $f \rightarrow 0$,
we get
\begin{equation}
      2D^{l}{\overline{J}}_{[ij][kl]}+\left(\frac{4}{3}
      {t_{k}}^{[lm]}-{\delta_{k}}^{[l}v^{m]}+{a_{k}}^{lm}\right)
      {\overline{J}}_{[ij][lm]}=0\;,
\end{equation}
where
\begin{equation}
      {\overline{J}}_{ijkl} \stackrel{\rm def}{=} 2{\overline{a}}_{3}R
                            \eta_{ik}\eta_{jl}+2\eta_{ik}
                            ({\overline{a}}_{1}E_{jl}+
                            {\overline{a}}_{2}I_{jl})\;.
\end{equation}
By substituting $R_{ijkl}=0$, which is a solution of (6.2), into (3.10),
we obtain
\begin{equation}
      -2D^{k}F_{ijk}+2v^{k}F_{ijk}+2H_{ij}-\eta_{ij}L_{T}
      =T_{ij}\;.
\end{equation}
Specifically, if we set ${A^{ij}}_{\mu }=0$, (6.4) with $\delta =0$
reduces to (3.7) of Ref. [11], which is the gravitational field
equation in a teleparallel theory. Even for this limiting case, the
theory given in this paper is not identical to the theory developed
in Ref. [11],
because (6.2) does not necessarily imply $R_{ijkl}=0$.
To put it briefly, the teleparallel theory developed in Ref. [11] is
\lq \lq included as a solution" in the limiting case with
$a_{i}\rightarrow \infty , (i=1,2,3)$ and with $\delta =0$.
\subsection{The case with $\alpha \rightarrow \infty, \beta
\rightarrow \infty, \gamma \rightarrow \infty$}
When $\alpha, \beta$ and $\gamma $ become large, the torsion tensor
becomes infinitely small with $H_{ijk}$ kept finite, which follows
from (3.16). In the limit of $\alpha \rightarrow \infty,
\beta \rightarrow \infty, \gamma \rightarrow \infty$, the underlying
space-time is of the Riemann type, and $H_{ijk}$ is given by
\begin{eqnarray}
     H_{ijk} & = & -2a_{2}\nabla_{[i}G_{j]k}(\{\})\nonumber \\
             & & -8\left(a_{3}+\frac{a_{2}}{6}\right)
             \eta_{k[i}\partial_{j]}G(\{\})-S_{ijk}\;,
\end{eqnarray}
which is obtained from (4.21). By using (2.8), (4.19) and (6.5) in
(4.15), we obtain
\widetext
\begin{eqnarray}
 & & 2aG_{ij}(\{\})+\frac{2}{3}(7a_{2}+6a_{3})R_{ij}(\{\})R(\{\})
     -8a_{2}{{R_{i}}^{k}}(\{\})R_{jk}(\{\}) \nonumber \\
 & & -\eta_{ij}\left(-3a_{2}R^{kl}(\{\})R_{kl}(\{\})
     +\frac{1}{3}(5a_{2}+3a_{3})(R(\{\}))^{2} \right)
     +2a_{2}\nabla^{k}\nabla_{k}G_{ij}(\{\}) \nonumber \\
 & & -\frac{4}{3}(a_{2}+6a_{3})(\eta_{ij}\nabla^{k}\nabla_{k}
     -\nabla_{i}\nabla_{j})G(\{\}) \nonumber \\
 & & =T_{ij}-\nabla^{k}(S_{ijk}-S_{ikj}-S_{jki})\;.
\end{eqnarray}
\narrowtext
\noindent This equation is obtainable also directly from the
Lagrangian $\tilde{L} \stackrel{\rm def}{=} L_{GR}
+L_{M}(\varphi , \nabla_{k}\varphi )$, where
\begin{eqnarray}
     L_{GR} \stackrel{\rm def}{=} a_{2}R^{kl}(\{\})R_{kl}(\{\})
             +\left(a_{3}-\frac{a_{2}}{3}\right)
             (R(\{\}))^{2} \nonumber \\
             +aR(\{\})\;,
\end{eqnarray}
\begin{equation}
     \nabla_{k}\varphi \stackrel{\rm def}{=} {e^{\mu }}_{k}
                        \left(\partial_{\mu }\varphi
                        +\frac{i}{2}{\Delta^{lm}}_{\mu }M_{lm}
                        \varphi \right)\;.
\end{equation}
\section{LINEARIZED GRAVITATIONAL FIELD EQUATIONS AND RELATION
TO THE NEWTON THEORY}
			    \setcounter{equation}{0}
\subsection{Linearized equations}
We now examine the gravitational field equations in the weak field
situations in which
\begin{equation}
     {a^{i}}_{\mu } \stackrel{\rm def}{=} {e^{i}}_{\mu }
     -{\delta^{i}}_{\mu }
\end{equation}
and ${A^{ij}}_{\mu }$ are so small that it is sufficient to keep
only terms linear in
${a^{i}}_{\mu }$ and in ${A^{ij}}_{\mu }$. In this approximation,
Greek and Latin indices need not be distinguished with each other,
and thus we shall use Greek indices throughout the present section
with the understanding that they are raised and lowered with
$(\eta^{\mu \nu }) \stackrel{\rm def}{=} {(\eta_{\mu \nu })}^{-1}$
and $(\eta_{\mu \nu }) \stackrel{\rm def}{=} diag(-1,1,1)$.
For this case, the components $g_{\mu \nu }$ of the metric tensor
has the expression
\begin{equation}
     g_{\mu \nu } = \eta_{\mu \nu }+h_{\mu \nu }
\end{equation}
with
\begin{equation}
     h_{\mu \nu } \stackrel{\rm def}{=} a_{\mu \nu}+a_{\nu \mu }\;.
\end{equation}
Noting (4.1) and (7.1), we employ $a_{\mu \nu }$ and the torsion
tensor $T_{\lambda \mu \nu }$ as independent field variables, and
express the linearized gravitational field equations in terms of
them. In what follows, we consider the case with $\delta =0$ only,
because the "cosmological term" $\delta $ in the Lagrangian $L_{T}$
is not in harmony with our weak field approximation.
After some calculations, we find that (4.15) and (4.21) take the forms
\begin{equation}
     2aG_{\mu \nu }^{(1)}-2\partial^{\lambda }
     {F'}_{\mu \nu \lambda}^{(1)} = T_{\mu \nu }
\end{equation}
and
\begin{equation}
      Z_{\lambda \mu \nu }^{(1)} = -S_{\lambda \mu \nu }\;,
\end{equation}
respectively. Here, we have defined
\widetext
\begin{equation}
      G_{\mu \nu}^{(1)} \stackrel{\rm def}{=}
                          \partial_{\lambda }\partial_{(\mu }
                          {h^{\lambda }}_{\nu )}-\frac{1}{2}
                          \partial_{\mu }\partial_{\nu }h
                          -\frac{1}{2}\Box h_{\mu \nu}
                          -\frac{1}{2}\eta_{\mu \nu }
                          (\partial_{\lambda }\partial_{\rho }
                          h^{\lambda \rho }-\Box h)\;,
\end{equation}
\begin{equation}
     {F'}_{\mu \nu \lambda }^{(1)}
         \stackrel{\rm def}{=}  \left(\alpha+\frac{2a}{3} \right)
         (t_{\mu \nu \lambda }-t_{\mu \lambda \nu })
         +\left(\beta -\frac{a}{2} \right)
         (\eta_{\mu \nu}v_{\lambda }-
         \eta_{\mu \lambda }v_{\nu }) +2\left(\gamma-\frac{a}{4}
         \right)a_{\mu \nu \lambda }=-{F'}_{\mu \lambda  \nu }^{(1)}\;,
\end{equation}
\begin{eqnarray}
     Z_{\lambda \mu \nu }^{(1)} &
     \stackrel{\rm def}{=} & a_{2}
      \left(\partial_{\lambda }G_{\mu \nu }^{(1)}-
      \partial_{\mu }G_{\lambda \nu }^{(1)}\right)
      +8\left(a_{3}+\frac{a_{2}}{6} \right)\eta_{\nu [\lambda }
      \partial_{\mu ]}G^{(1)}
      +(a_{1}+a_{2})\eta_{\nu [\mu }\partial_{\rho }
      \partial_{\sigma }{t^{\rho \sigma }}_{\lambda ]}
      \nonumber \\
  & & +\frac{1}{3}[\partial_{\mu }\partial_{\sigma }\{2a_{1}
      {t^{\sigma }}_{[\lambda \nu ] }
      +3a_{2}{t_{\lambda \nu }}^{\sigma }\}
      -\partial_{\lambda }\partial_{\sigma }
      \{2a_{1}{t^{\sigma }}_{[\mu \nu ]}
      +3a_{2}{t_{\mu \nu }}^{\sigma }\}] \nonumber \\
  & & +\frac{1}{6}(3a_{1}+a_{2}-48a_{3})
      \eta_{\nu [\mu }\partial_{\lambda ]}
      \partial_{\sigma }v^{\sigma }
      +\frac{1}{2}(a_{1}+a_{2})\{\partial_{\nu }\partial_{[\mu }
      v_{\lambda ]}+\eta_{\nu [\lambda }\Box v_{\mu ]}\} \nonumber \\
  & & +a_{1}\partial_{\sigma }\partial_{[\mu }
      {a_{\lambda ]\nu }}^{\sigma }+H_{\lambda \mu \nu }^{(1)}
\end{eqnarray}
with $G^{(1)} \stackrel{\rm def}{=}
\eta^{\mu \nu }G_{\mu \nu }^{(1)}$, $h \stackrel{\rm def}{=}
\eta^{\mu \nu }h_{\mu \nu }$, $\Box \stackrel{\rm def}{=}
\partial^{\mu }\partial_{\mu }$ and
\begin{equation}
     H_{\lambda \mu \nu }^{(1)} \stackrel{\rm def}{=}
                                 -\left(\alpha +\frac{2a}{3}\right)
                                 (t_{\nu \lambda \mu }-
                                 t_{\nu \mu \lambda })
                                 -\left(\beta -\frac{a}{2} \right)
                                 (\eta_{\nu \lambda }v_{\mu }
                                 -\eta_{\nu \mu }
                                 v_{\lambda })+(4\gamma  -a)
                                 a_{\lambda \mu \nu }
                                 =-H_{\mu \lambda \nu }^{(1)}\;.
\end{equation}
\narrowtext
\noindent These $G_{\mu \nu }^{(1)}$,
${F'}_{\mu \nu \lambda }^{(1)}$, $Z_{\lambda \mu \nu }^{(1)}$,
$G^{(1)}$ and $H_{\lambda \mu \nu }^{(1)}$ are the linearized
expressions for $G_{\mu \nu }(\{\})$, $F'_{\mu \nu \lambda }$,
$Z_{\lambda \mu \nu }$, $G(\{\})$ and $H_{\lambda \mu \nu }$,
respectively. In the lowest order approximation now considering,
we have the differential conservation law,
\begin{equation}
     \partial_{\nu }{T_{\mu }}^{\nu } = 0
\end{equation}
and the Tetrode formula,
\begin{equation}
     \partial_{\lambda}{S_{\mu \nu }}^{\lambda } = T_{[\mu \nu]}\;.
\end{equation}
The field equations (7.4) and (7.5) are invariant under the
infinitesimal gauge transformations:
\begin{equation}
     h^{*}_{\mu \nu}=h_{\mu \nu }+\partial_{\mu }{\Lambda }_{\nu }
           +\partial_{\nu }{\Lambda }_{\mu }\;,
\end{equation}
\begin{equation}
     a^{*}_{[\mu \nu ]}=a_{[\mu \nu ]}+\omega_{\mu \nu }\;, \;
         \omega_{\mu \nu }=-\omega_{\nu \mu }
\end{equation}
with $\Lambda_{\mu }$ and $\omega_{\mu \nu }$
being both arbitrary infinitesimal functions. The transformations
(7.12) and (7.13) are the infinitesimal versions of the general
coordinate and local gauge transformations, respectively.
The invariance under the transformation (7.12) means that the
antisymmetric part $a_{[\mu \nu ]}$ does not have physical
significance. By virtue of the invariance under the transformation
(7.12), we can put the harmonic condition,
\begin{equation}
     \partial^{\nu }{\overline{h}}_{\mu \nu }=0\;,
\end{equation}
which is assumed from now on. Here, we have defined
\begin{equation}
    {\overline{h}}_{\mu \nu } \stackrel{\rm def}{=} h_{\mu \nu }
                              -\frac{1}{2}\eta_{\mu \nu }h\;.
\end{equation}
The linearized Einstein tensor now takes the form,
\begin{equation}
     G_{\mu \nu }^{(1)}=-\frac{1}{2}\Box \overline {h}_{\mu \nu }\;.
\end{equation}
It is convenient to decompose (7.4) into the symmetric and
antisymmetric parts:
\begin{eqnarray}
 & & 2aG_{\mu \nu }^{(1)}-3\left(\alpha +\frac{2a}{3}\right)
     \partial^{\lambda }t_{\mu \nu \lambda } \nonumber \\
 & & -2\left(\beta-\frac{a}{2} \right )(\eta_{\mu \nu }
     \partial^{\lambda }v_{\lambda }-\partial_{(\mu }v_{\nu )})
     = T_{(\mu \nu )}\;,
\end{eqnarray}
\begin{eqnarray}
     2\left(\alpha +\frac{2a}{3} \right)\partial^{\lambda }
     t_{\lambda [\mu \nu ]}
     +2\left(\beta -\frac{a}{2} \right)\partial_{[\mu }v_{\nu ]}
     \nonumber \\
     -4\left(\gamma -\frac{a}{4} \right)
     \partial^{\lambda }a_{\mu \nu \lambda } = T_{[\mu \nu ]}\;.
\end{eqnarray}
Taking the trace of (7.17), we obtain
\begin{equation}
     2aG^{(1)}-4\left(\beta -\frac{a}{2}\right )\partial^{\lambda }
     v_{\lambda } = T
\end{equation}
with $T \stackrel{\rm def}{=} \eta^{\mu \nu }T_{\mu \nu }$.
Both sides of (7.4) are divergenceless because of (7.10), while the
divergence of (7.5) with respect to ${\nu }$ gives (7.18) by virtue
of (7.11). Thus, the field equations (7.4) and (7.5) give
$(9+9)-(3+3)=12$ independent equations for $3+9=12$ independent
field variables.
\subsection{$h_{00}$ due to a static, \lq \lq spin"less point
like source}
Using (7.8), (7.17) and (7.19) in the the symmetric part of the
divergence of (7.5) with respect to $x^{\lambda }$,
\begin{equation}
    \partial^{\lambda }Z_{\lambda (\mu \nu )}^{(1)}
    = -\partial^{\lambda }S_{\lambda (\mu \nu )}\;,
\end{equation}
we obtain the fourth-order field equation for
${\overline{h}}_{\mu \nu }$,
\begin{equation}
    A\Box {\overline{h}}_{\mu \nu }
    +B\Box^{2}{\overline{h}}_{\mu \nu }
    +C(\eta_{\mu \nu }\Box -\partial_{\mu }\partial_{\nu })\Box
    \overline{h} = T_{\; \, \mu \nu }^{(\rm eff )}
\end{equation}
with
\begin{equation}
    A \stackrel{\rm def}{=} -a\;,
\end{equation}
\begin{equation}
    B \stackrel{\rm def}{=}
    -\frac{3\alpha a_{2}}{3\alpha+2a}\;,
\end{equation}
\begin{equation}
    C \stackrel{\rm def}{=} \frac{1}{6(2\beta -a)}
      \{8\beta (a_{2}+6a_{3})-3aa_{2} \}
      -\frac{aa_{2}}{3\alpha +2a}
\end{equation}
and
\begin{eqnarray}
    T_{\; \, \mu \nu }^{(\rm eff )}
    & \stackrel{\rm def}{=} & T_{(\mu \nu)}-2\partial^{\lambda }
     S_{\lambda (\mu \nu )}-\frac{a_{2}+24a_{3}}{6(2\beta -a)}
     (\eta_{\mu \nu }\Box -\partial_{\mu }\partial_{\nu })T
     \nonumber \\
 & & -\frac{2a_{2}}{3\alpha +2a}\left(\Box T_{(\mu \nu )}
     -\frac{1}{2}(\eta_{\mu \nu }\Box
     -\partial_{\mu }\partial_{\nu })T \right. \nonumber \\
 & & +2\partial^{\lambda }\partial^{\rho }\partial_{(\mu }
     S_{\nu )\lambda \rho } \biggr)\;.
\end{eqnarray}
It is worth mentioning that the parameters $a_{1}$ and $\gamma $ do
not appear in (7.21) with (7.22)--(7.25).
We consider now the gravitational field produced by a static,
\lq \lq spin"less source located at the origin, for which
$S_{\lambda \mu \nu }$ is vanishing and $T_{\mu \nu }$ is given by
\begin{equation}
     T_{\mu \nu }=\left\{ \begin{array}{ll}
                              Mc^{2}\delta^{2}(\vec{r})\;,
                               \mbox{$\mu =\nu =0\;,$} \\
                              0\; \; \; \; \; \; \; \; \;
                               \; \; \; \; \;,
                              \mbox{otherwise}
                              \end{array}\right.
\end{equation}
with $\vec{r} \stackrel{\rm def}{=} (x^{1},x^{2})$.
For this case, $T_{\; \, \mu \nu }^{(\rm eff )}$ has the expression
\begin{equation}
\left.
\begin{array}{l}
     T_{\; \: 0 0}^{(\rm eff)}=Mc^{2}\left\{\delta^{2}(\vec{r})
     +(P+Q)\Delta \delta^{2}(\vec{r}) \right\}\;, \\
     T_{\; \: 0 \alpha }^{(\rm eff )}=
             T_{\; \: \alpha 0}^{(\rm eff )}=0\;, \\
     T_{\; \: \alpha \beta }^{(\rm eff )}=Mc^{2}(P-Q)
                             (\partial_{\alpha }\partial_{\beta }
                             -\delta_{\alpha \beta}\Delta )
                             \delta^{2}(\vec{r})\;,
\end{array}
\right\}
\end{equation}
where the indices $\alpha $ and $\beta $
run over 1 and 2,   \\
$\Delta \stackrel{\rm def}{=} (\partial_{1})^{2}
+(\partial_{2})^{2}$ and
\begin{equation}
      P \stackrel{\rm def}{=} -\frac{a_{2}+24a_{3}}{6(2\beta -a)}\;,
      \: \: Q \stackrel{\rm def}{=} -\frac{a_{2}}{3\alpha +2a}\;.
\end{equation}
Taking trace of (7.21), we obtain
\begin{equation}
      A\Box \overline{h}+(B+2C)\Box^{2}\overline{h} =
      T^{(\rm eff )}\;,
\end{equation}
where $\overline{h} \stackrel{\rm def}{=} \eta^{\mu \nu }
{\overline{h}}_{\mu \nu }$, and
\begin{equation}
      T^{(\rm eff )} \stackrel{\rm def}{=} \eta^{\mu \nu }
                      T_{\; \, \mu \nu }^{(\rm eff )}
                     =-Mc^{2}\delta^{2}(\vec{r})-2Mc^{2}
                      P\Delta \delta^{2}(\vec{r})\;.
\end{equation}
In the following, we shall solve (7.21) and (7.29) to give a
time-independent circularly symmetric potential $h_{00}$ for each of
four cases.
\subsubsection{The case with $a \neq 0$}
For this case, (7.29) is solved, by utilizing the method of Fourier
integral, to give
\widetext
\begin{eqnarray}
   \overline{h}(r) & = & -\frac{Mc^2}{2\pi A}\log r+
                    \frac{Mc^2}{4\pi^2 A}
                    (2PA-B-2C)\int \frac{e^{i\vec{k} \cdot \vec{r}}}
                    {(B+2C){\vec{k}}^{2}-A}d^{2}\vec{k} \nonumber \\
               & &  +C_{1}\int \delta ((B+2C){\vec{k}}^{2}-A)
                    e^{i\vec{k} \cdot \vec{r}}d^{2}\vec{k}+C_{2}
\end{eqnarray}
\narrowtext
\noindent with $C_{1}$ and $C_{2}$ being integration constants,
$r \stackrel{\rm def}{=} |\vec{r}|$ and
$\vec{k} \stackrel{\rm def}{=} (k_{1}, k_{2})$. Substituting (7.31)
into (7.21) with $\mu =\nu =0$, solving the equation
thus obtained and using (7.22), (7.23) and (7.24) and the formula,
\begin{equation}
    J_{0}(kr)=\frac{1}{2\pi }\int_{0}^{2\pi }e^{ikr\cos \theta }
              d\theta \;,
\end{equation}
we obtain
\widetext
\begin{eqnarray}
h_{00}(r) & = & -\frac{Mc^{2}a_{2}(3\alpha +2a)}{4\pi a}
          \int_{0}^{\infty } \frac{kJ_{0}(kr)}{3\alpha a_{2}
          k^{2}-a(3\alpha +2a)}dk \nonumber \\
     & &  +\frac{Mc^{2}(a_{2}+24a_{3})(2\beta -a)}{4\pi a}
          \int_{0}^{\infty } \frac{kJ_{0}(kr)}
          {2\beta (a_{2}+24a_{3})k^{2}
          +3a(2\beta -a)}dk \nonumber \\
     & &  +6\pi C_{1}|2\beta -a|\int_{0}^{\infty }
          \delta (2\beta (a_{2}+24a_{3})k^{2}
          +3a(2\beta -a))kJ_{0}(kr)dk \nonumber \\
     & &  +C_{3}|3\alpha +2a|
          \int_{0}^{\infty }\frac{\delta (3\alpha a_{2}k^{2}
          -a(3\alpha +2a))J_{0}(kr)}{k}dk+C_{4}
\end{eqnarray}
\narrowtext
\noindent with $k \stackrel{\rm def}{=} |\vec{k}|$, and $C_{3}$ and
$C_{4}$ being integration constants. Here, $J_{0}$ denotes
the Bessel function of the first kind and of index zero.
\subsubsection{The case with $a=0$ and
$\alpha \beta a_{2}(a_{2}+24a_{3}) \neq 0$}
In a way similar to the case {\em 1}, the solution
\begin{eqnarray}
h_{00}(r) & = & \frac{Mc^{2}(3\alpha +4\beta )}
           {24\pi \alpha \beta }\log r-\frac{Mc^{2}(a_{2}+6a_{3})}
           {4\pi a_{2}(a_{2}+24a_{3})}r^{2}\log r \nonumber \\
          & & +C_{5}r^{2}+C_{6}
\end{eqnarray}
is obtained, where $C_{5}$ and $C_{6}$ are integration constants.
\subsubsection{The case with $a_{2} \rightarrow \infty $,
   $a_{3} \rightarrow \infty $ and $\alpha \beta \neq 0$}
The parameters $a_{2}$ and $a_{3}$ are assumed to have the expression
\begin{equation}
   a_{i}=\frac{{\overline{a}}_{i}}{f^{2}},\;(i=2,3)
\end{equation}
with $f$ being a real constant, and we consider the limiting case
with $f \rightarrow 0$. Multiplying both sides of (7.21) and (7.29)
by $f^{2}$ and taking the limit $f \rightarrow 0$ and following
a similar procedure as in the above, we find
\begin{equation}
  h_{00}(r) = \frac{3\alpha +4\beta }{24\pi \alpha \beta }
             Mc^{2}\log r+C_{7}r^{2}+C_{8}
\end{equation}
with $C_{7}$ and $C_{8}$ being integration constants.
\subsubsection{The case with $\alpha \rightarrow \infty $,
$\beta \rightarrow \infty $ and $a \neq 0$}
The potential for this case is given by
\widetext
\begin{eqnarray}
h_{00}(r) &=& -\frac{Mc^{2}a_{2}}{4\pi a}
          \int_{0}^{\infty }\frac{kJ_{0}(kr)}{a_{2}k^{2}-a}dk
          +\frac{Mc^{2}(a_{2}+24a_{3})}{4\pi a}
          \int_{0}^{\infty }\frac{kJ_{0}(kr)}
          {(a_{2}+24a_{3})k^{2}+3a}dk \nonumber \\
     & &  +\frac{2C_{9}(a_{2}+6a_{3})}{3}
          \int_{0}^{\infty }\frac{\delta ((a_{2}+24a_{3})k^{2}+3a)
          kJ_{0}(kr)}{a_{2}k^{2}-a}dk \nonumber \\
     & &  +C_{9}\int_{0}^{\infty }\frac{\delta
          ((a_{2}+24a_{3})k^{2}+3a)J_{0}(kr)}
          {k}dk+C_{10}\int_{0}^{\infty } \frac{\delta (a_{2}k^{2}-a)
          J_{0}(kr)}{k}dk+C_{11}
\end{eqnarray}
\narrowtext
\noindent with $C_{i}$ ($i$=9, 10, 11) being integration
constants.
\subsection{Relation to the Newton theory}
We consider a macroscopic body for which effects due to the
\lq \lq spin"s of the fundamental constituent particles can be
ignored. The world line of motion of this macroscopic body moving
freely in the space-time under consideration is the geodesic line of
the metric $g_{\mu \nu }dx^{\mu }\otimes d^{\nu }$, as has been known
in Sec.V. Thus, when the motion is sufficiently slow and the
gravitational field is weak, this body obeys the equation of motion,
\begin{equation}
      \frac{d^{2}\vec{r}}{dt^{2}}=-\frac{\partial U}{\partial \vec{r}}
\end{equation}
with $U \stackrel{\rm def}{=} -c^{2}h_{00}/2$.
Neither of the solutions (7.33) and (7.37) can give a Newton potential.
The potential $U$ given by the solution (7.34) satisfies the Newton
equation for the gravitational potential,
\begin{equation}
    \Delta U = 4\pi GM\delta (\vec{r})\;,
\end{equation}
if the conditions
\begin{equation}
          3\alpha +4\beta =-\frac{96\alpha \beta \pi G}{c^{4}}\;,
\end{equation}
\begin{equation}
       a_{2}+6a_{3}=0\;,
\end{equation}
are both satisfied {\em and when the integration constants $C_{5}$
is chosen to be zero}: $C_{5}=0$. Here, $G$
stands for \lq \lq Newton gravitational constant".
Also the solution (7.36) gives the Newton potential, if the
condition (7.40) is satisfied {\em and if} $C_{7}=0$.
Thus, each of the case {\em 2} with the conditions (7.40) and (7.41)
and of the case {\em 3} with the condition (7.40) can give a Newtonian
limit by a suitable choice of the integration constant. But, it
should be noted that {\em the field equations for $h_{\mu \nu }$ are
fourth-order differential equations for both cases}.
\section{SUMMARY AND COMMENTS}
                              \setcounter{equation}{0}
We have formulated a Poincar\'{e} gauge theory of (2+1)-dimensional
gravity and the results can be summarized as follows:

(1) The theory is underlain with the Riemann-Cartan space-time, and
the gravity is attributed to the curvature and the torsion. The most
general gravitational Lagrangian, which is at most quadratic in
curvature and torsion tensors, is given by $L_{G}
\stackrel{\rm def}{=} L_{T}+L_{R}$ with $L_{T}$ and $L_{R}$ being
given by (3.1) and (3.2), respectively.

(2) The gravitational field equations are given by (3.10) and (3.16),
the alternative forms of which are (4.15) and (4.21),
respectively.

(3) Solutions of the vacuum Einstein equation with the cosmological
constant $\Lambda$ satisfy the vacuum gravitational field
equations with $\delta =2\Lambda (a+6a_{3}\Lambda)$.

(4) Equation (4.21) does not contain third derivatives of the metric
tensor, if and only if $a_{2}=a_{3}=0$. For the case with
$a_{2}=a_{3}=0$, the vanishing torsion satisfies (4.21) with
$S_{ijk} \equiv 0$, and (4.15) reduces to the Einstein equation (4.25)
for the vanishing torsion.

(5) For the case with $a_{1}=a_{2}=a_{3}=0$ and with
$(3\alpha +2a)(2\beta -a)(4\gamma -a)\neq 0$, the torsion is
\lq \lq frozen" at the place where the \lq \lq spin" density
$S_{ijk}$ does not vanish. If $a=1/2\kappa $ in addition and
the source field is \lq \lq spin"less, field equations
for the gravitational and source fields agree with those in the
Einstein theory.

(6) The world line of the macroscopic body is the geodesic line of the
metric $g_{\mu \nu }dx^{\mu }\otimes dx^{\nu }$, if the effects due
to the \lq \lq spin" of the fundamental constituent particles can be
ignored.

(7) In the sense mentioned in Sec. VI A, the teleparallel theory
developed in Ref. [11] is \lq \lq included as a solution" in the
limiting case with $a_{i}\rightarrow \infty (i=1,2,3)$ and with
$\delta =0$.

(8) For the case with $\alpha \rightarrow \infty ,
\beta \rightarrow \infty ,\gamma \rightarrow \infty $, the
underlying space-time is of the Riemann type.

(9) The linearized field equations lead to the fourth-order
differential equation (7.21) for weak gravitational potentials. For
the gravitational field produced by a static \lq \lq spin"less point
like source, (7.21) has been solved to give the potential
$U \stackrel{\rm def}{=} -c^{2}h_{00}/2$. The solutions are
classified into the four cases {\em 1}, {\em 2}, {\em 3} and
{\em 4}. Each of the case {\em 2} with the conditions (7.40) and
(7.41) and of the case {\em 3} with the condition (7.40)
can give a Newtonian limit by a suitable choice of the
integration constant.

The following is worth to be mentioned:

(a) Even for the case with $a_{1}=a_{2}=a_{3}=0, a=1/2\kappa $
and $(3\alpha +2a)(2\beta -a)(4\gamma -a)\neq 0$,
field equations are different from those in the Einstein theory,
if the source field has non-vanishing \lq \lq spin". For this case,
space-times in the vacuum regions are locally the same as those
in the Einstein theory, but the quantized theory and non-local
properties such as the gravitational Aharonov-Bohm effect
[17, 18] due to the \lq \lq spin"ing source fields are presumably
different from those of the latter theory.

(b) The condition (7.40) agrees with the condition (5.11) in Ref. [11]
in its form. This is quite naturally understood, if we note the
discussions in Sec.VI A and in Sec.VII B {\em 3} and the fact that
the parameter $a_{1}$ does not appear in the linearized equation
(7.21) with (7.22)--(7.25).

(c)As is known from (3), black hole solutions [19, 20] of the
ordinary three-dimensional vacuum Einstein equation with a
negative cosmological constant satisfy the vacuum gravitational
field equations in our theory. Also, it is worth adding that these
solutions are independent of the black holes in a teleparallel
theory discussed in Ref. [11].
\begin{center}
                          {\bf ACKNOWLEDGMENTS}
\end{center}

The author would like to express his sincere thanks to K. Takashiba,
A. Ni\'{e}gawa and N. Asida for giving helpful instructions on the
\TeX-typing.


\begin{references}
\bibitem[12]{LAG}From now on, Lagrangian density is simply called
 Lagrangian.
\bibitem[15]{ASYM}We denote symmetrization and antisymmetrization of
 tensor indices by a round bracket ( ) and a square bracket [ ],
 respectively: For example, $J_{[ik][jl]} \stackrel{\rm def}{=}
 (1/4)(J_{ikjl}-J_{kijl}-J_{iklj}+J_{kilj})$.
\bibitem[16]{SPN}By \lq \lq spin", we mean the quantum number
 associated with the three-dimensional Lorentz group.
\end{references}
\end{document}